\documentclass[letterpaper,10pt]{article}
\pdfoutput=1
\usepackage[utf8]{inputenc} 
\usepackage{tabularx} 
\usepackage{dsfont}
\usepackage{amsmath,amssymb,amsfonts,amsthm,amssymb}   
\usepackage{calc} 
\usepackage{xcolor}
\usepackage{listings} 
\usepackage{cite} 
\usepackage{graphicx} 
\usepackage[margin=1in,letterpaper]{geometry} 

\usepackage[title]{appendix}

\usepackage{wasysym} 

\usepackage[
 colorlinks=true,       
    linkcolor=blue,        
    citecolor=blue,        
    filecolor=magenta,     
    urlcolor=blue 
      pdfstartview=FitV,
      pdftitle={},
        pdfauthor={David Gordo},
        pdfsubject={},
        pdfkeywords={},
        pdfpagemode=None,
        bookmarksopen=true    
      ]{hyperref}
\usepackage{multicol} 
\usepackage{multirow}
\usepackage{xcolor} 
\usepackage{makeidx} 
\usepackage{ccaption} 

\catcode`\@=11 

\numberwithin{equation}{section}

\def\ie{{\it i.e.}}

\def\revise#1       {\raisebox{-0em}{\rule{3pt}{1em}}%
                     \marginpar{\raisebox{.5em}{\vrule width3pt\
                     \vrule width0pt height 0pt depth0.5em
                     \hbox to 0cm{\hspace{0cm}{%
                     \parbox[t]{4em}{\raggedright\footnotesize{#1}}}\hss}}}}

\def\calc         {{\cal C}}

\def\calk         {{\cal K}}

\def\caln         {{\cal N}}
\def\calo         {{\cal O}}

\def\calw         {{\cal W}}

\def\del          {\partial}

\def\de#1#2{{\rm d}^{#1}  #2 \, }

\def\sqr#1#2{{\vcenter{\vbox{\hrule height.#2pt
 \hbox{\vrule width.#2pt height#1pt \kern#1pt
 \vrule width.#2pt}\hrule height.#2pt}}}}

\def\aa1{\phi}
\def\cc1{\psi}

\newcommand{\abs}[1]{\left| #1 \right|} 
\newcommand{\norm}[1]{\lvert #1 \rvert} 
\newcommand{\ket}[1]{\left| #1 \right>} 
\newcommand{\bra}[1]{\left< #1 \right|} 
\newcommand{\braket}[2]{\left\langle #1 \vphantom{#2} \right|
 \left. #2 \vphantom{#1} \right\rangle} 
 \newcommand{\matrixel}[3]{\left\langle #1 \vphantom{#2#3} \right|
 #2 \left| #3 \vphantom{#1#2} \right\rangle} 
 \newcommand{\stringa}{\ttfamily\lstinline}
\def\cod#1{{\stringa!#1!}}

 \def\bam{\bar{\alpha}_{\mu_R}}
 \def\ido {\mathds{1}}

\newcommand\plotall[2]{
\begin{figure}[h]
\begin{center}$
\begin{array}{cc}
\includegraphics[width=8cm]{images/R21_b_Y#1.png} & \includegraphics[width=8cm]{images/R32_b_Y#1.png} 
\\ 
\includegraphics[width=8cm]{images/R10_b_Y#1.png} & \includegraphics[width=8cm]{images/R20_b_Y#1.png}
\end{array}$
\includegraphics[width=8cm]{images/R30_b_Y#1.png}
\end{center}
\caption{Functional dependence of the different $R^n_m$ functions on the veto $b$ at $Y=\,\,$#2. The reddish band denotes the experimental uncertainty 
while the blueish band denotes the theoretical uncertainty from the renormalisation scale variation.}
\end{figure}
}

\begin{document}

\title{Inclusive dijet hadroproduction with a rapidity veto constraint}
\author{F. Caporale, F.~G. Celiberto, G. Chachamis, \\ 
        D. Gordo G{\' o}mez,  A. Sabio Vera\\ \\
{\small Instituto de F{\' \i}sica Te{\' o}rica UAM/CSIC, Nicol{\'a}s Cabrera 15}\\ 
{\small \& Universidad Aut{\' o}noma de Madrid, E-28049 Madrid, Spain.}
}

\date{\today}
\maketitle

\begin{abstract}
We study ratios of azimuthal-angle distributions in Mueller-Navelet jets after imposing a rapidity veto constraint: the minijet radiation activity is restricted to only allow final-state partons separated at least a distance in rapidity $b$. It is well-known that the asymptotic growth with the rapidity separation of the two tagged jets of the NLLA BFKL Green's function requires a value of $b \simeq {\cal O} (2)$ in order to avoid unphysical cross sections. We further investigate this point from a phenomenological point of view and work out those values of $b$ which best fit angular distributions measured at the LHC in a realistic set-up where impact factors and parton distribution effects are also taken into account. 
\end{abstract}

\section{Introduction}

The phenomenological study of processes at the high energy limit of Quantum Chromodynamics (QCD) offers invaluable information that goes beyond the perturbative regime of the theory. It touches important issues such as factorization between soft and hard physics and it questions the validity of fixed order calculations. At asymptotically high energies, the convergence of the perturbative expansion that is truncated at a certain order in the strong coupling $\alpha_s$ is not \textit{a priori} guaranteed. This is because large logarithms in the scattering center-of-mass energy  squared, $(\log  s)^n$,
appear in Feynman diagrams to all orders (arbitrarily high order $n$ in the expansion) and one needs to make sure that they are resummed properly. 
A powerful approach to perform the resummation is the Balitsky-Fadin-Kuraev-Lipatov (BFKL) framework initially developed at leading logarithmic approximation 
(LLA)~\cite{Balitsky:1978ic,Kuraev:1977fs,Kuraev:1976ge,Lipatov:1976zz,Fadin:1975cb,Lipatov:1985uk}, where terms of the form $(\alpha_s \log(s))^n$ were resummed. In order to improve the LLA accuracy, the next-to-leading logarithmic approximation (NLLA) corrections to the BFKL kernel were calculated~\cite{Fadin:1998py,Ciafaloni:1998gs}, where also terms that behave like $\alpha_s (\alpha_s \log s)^n$ were taken into account.  It was seen however,  that at NNLA the positiveness of cross sections was not always ensured. This is due to the presence of large collinear logarithms that need extra treatment, a step that led to the so-called collinearly improved BFKL kernel~\cite{Salam:1998tj,Ciafaloni:2003kd} allowing for more robust phenomenological studies based on NLLA BFKL. Obviously, an important question for collider phenomenology is gauging reliably at which energies  the BFKL dynamics becomes relevant and cannot be ignored. 

At hadronic colliders and in particular at the LHC, a phenomenologically interesting process is Mueller-Navelet (MN) jet production~\cite{Mueller:1986ey}. MN jets (dijet production)  are inclusive final states where two jets with transverse momenta of similar sizes, $k_{A,B}$ are tagged to have a large rapidity separation $Y$. The presence of two hard but similar in size scales ($k_A$ and $k_B$) ensures in principle the applicability of a BFKL-based approach.
A number of works in the literature~\cite{DelDuca:1993mn,Stirling:1994he,Orr:1997im,Kwiecinski:2001nh,Andersen:2006pg,DeRoeck:2009id,Angioni:2011wj,Caporale:2013uva, Caporale:2012ih,Marquet:2007xx,Colferai:2010wu,Ducloue:2013wmi,Ducloue:2014koa,Mueller:2015ael,Chachamis:2015crx,N.Cartiglia:2015gve}  focuses on the azimuthal angle ($\theta$) behaviour of the two jets. This behavior is driven by the presence of decisive minijet activity in the rapidity space between the two outermost jets which in BFKL based studies is accounted for by a BFKL gluon Green's function connecting the two jets, $\varphi\left(k_A, k_B,Y \right)$.
It was shown by Schwennsen and one of us~\cite{Vera:2006un,Vera:2007kn}, that ratios of projections on azimuthal angle observables
\begin{eqnarray}
{\cal R}^m_n = \langle \cos{(m \, \theta)} \rangle / \langle \cos{(n \, \theta)} \rangle\,\,,
\label{gcr}
\end{eqnarray}
(where $m,n$ are integers) are the more favourable quantities in the search for a clear signal of BFKL effects. The comparison of different NLLA
calculations for these ratios ${\cal R}^m_n$~\cite{Ducloue:2013bva,Caporale:2014gpa,Caporale:2015uva,Celiberto:2015yba,Celiberto:2016ygs} against LHC experimental data has been promising so far.

Recently, a generalization of the azimuthal ratios in Eq.~(\ref{gcr}) was proposed for processes that have three~\cite{Caporale:2015vya,Caporale:2016soq, Caporale:2016zkc}
and four final state jets~\cite{Caporale:2015int,Caporale:2016xku}. These can be seen as special MN cases since 
the outermost jets still need to have a large rapidity distance and any other tagged jet is to be found in more central regions of the detector. 
In Ref.~\cite{Caporale:2015vya}, it has been argued that the new observables present additional advantages to the MN jets in the effort to disentangle a BFKL signal. However, there are no experimental analyses to compare to yet and hereafter we will restrict our discussion to MN jets.

In Refs.~\cite{CMS:2013eda,Khachatryan:2016udy} one can find a comparison between experimental data and theoretical predictions
for a number of MN azimuthal ratios. The theoretical predictions are obtained from the usual collinear Monte Carlo tools and from a BFKL based approach. The latter is to NLLA accuracy computed in the so-called Brodsky-Lepage-Mackenzie (BLM) scheme~\cite{Brodsky:1982gc}. It turns out that working in the BLM scheme is only essential in having a good description of the data when one studies ${\cal R}^m_n$ with either $m=0$ or $n=0$.
 
As we mentioned previously, higher order corrections to the BFKL equation (corrections beyond the LLA) are very important for both theoretical and phenomenological studies of QCD at high energies. So far, there is no unique approach on how to properly account for the corrections beyond the NLLA and for every process. It is known, however,  that the largest portion of the NLLA corrections are due to effects related to the running of 
$\alpha_s$ and to collinear contributions, while it has been argued  that the NLLA kernel induces large and negative logarithms in the ratio $k_A/k_B$ at NNLLA and beyond. Although these logarithms lie formally beyond the boundary of the NLLA approach, they can induce spurious large effects. For consistency with the DGLAP approach we know that they must be resummed to all orders. In Ref.~\cite{Salam:1998tj} it was shown how to extend the NLLA kernel so as to guarantee exactly that resummation, however, the prescription for modifying the kernel is not unique.
Schmidt, in Ref.~\cite{Schmidt:1999mz}, pointed out that a significant reduction in the resultant Green's function occurs if one only considers diagrams in which emitted gluons (minijets) have a minimum rapidity distance, $b$, relative to the preceding emitted gluon or, in other words, if one imposes a rapidity veto\footnote{The original presentation of this idea was by L.~N.~Lipatov at a talk presented at the 4th Workshop on Small-x and Diffractive
Physics, Fermi National Accelerator Laboratory, Sept. 17-20, 1998.} on the minimal rapidity distance between two subsequent minijet emissions. Furthermore, in~\cite{Forshaw:1999xm} it was shown that the large effect of imposing such a restriction accounts for the same regions of phase space in the NLLA corrections to the BFKL Green's function as the  collinear summation as proposed in~\cite{Salam:1998tj}. Recently~\cite{Ross:2016kzz}, the effect of a rapidity veto on the discrete BFKL Pomeron was studied whereas there were works in which a rapidity veto was applied to non linear evolution equations~\cite{Chachamis:2004ab,Enberg:2006aq}.

In this paper, we want to address the following questions: is it possible to obtain a good theoretical description of the ratios ${\cal R}^m_n$ including the ones with either $m=0$ or $n=0$ without necessarily using the BLM scheme? Is it possible to achieve that by employing a single global scale such as the rapidity veto? If indeed employing a rapidity veto allows in principle for a fit of the data, at what values of $b$ this happens?  Do the optimal $b$ values tell us how far from asymptotia we are at LHC energies? Here, we remind the reader of
a key conclusion from previous studies~\cite{Forshaw:1999xm} which was that the rapidity veto samples the region of phase space corresponding to collinear emissions already at a typical value that is somewhat larger than two units of rapidity for very high colliding energies, that is, well into the asymptotia region.

The structure of this paper is as follows. In Section 2 we introduce the necessary notation and conventions for the description of MN jets, important formulas for the NLLA gluon Green's function and finally the notion of the rapidity veto and the definition of the azimuthal angle correlations and their ratios. In Section 3, we study numerically the functional dependence of the ratios  ${\cal R}^m_n$ on the rapidity veto $b$ after imposing the same kinematical cuts as the ones used for the experimental analysis in Refs.~\cite{CMS:2013eda,Khachatryan:2016udy}. Finally, we present our conclusions in Section 4.

\section{Notation and Conventions}

MN jets was proposed by Mueller and Navelet~\cite{Mueller:1986ey} as a process in hadron colliders for which one could disentangle the high-energy behavior of the partonic cross section after removing most of  the parton distribution functions (PDFs) dependence. In collisions of two hadrons  (protons at the LHC) MN jets is the final state characterized by two tagged jets well separated in rapidity
\begin{equation} \label{eq:reaction}
p\left(p_A \right) + p\left(p_B\right) \rightarrow J_A \left(k_A\right) + J_B  \left(k_B \right) + X\, .
\end{equation}
The relevant kinematical configuration is given by\footnote{All transverse two-momenta will be denoted by vector variables.}
\begin{equation}\label{eq:scales}
s \equiv \left(p_A + p_B \right)^2 \gg Q^2 \sim \vec{k}_A^2 \sim  \vec{k}_B^2 \gg \Lambda_{QCD}^2
\end{equation}
where $Q$ is some typical hard transverse scale that guarantees the applicability of perturbation theory.  Using Sudakov decomposition with $p_{A,B}$ as a basis\footnote{The mass of the jets is neglected.}, we have
\begin{equation}\label{eq:SudDecomposition}
k_A=x_{J_A} p_A + \frac{\vec{k}_A^2}{x_{J_A} s} p_B + k_{A,\perp}\, \, , \, \,
k_B=x_{J_B} p_B + \frac{\vec{k}_B^2}{x_{J_B} s} p_A + k_{B,\perp} \, \, , \, \,  k_{A,B,\perp}^2=-\vec{k}_{A,B}^2\, ,
\end{equation}
where $x_{J_{A,B}}$ are the longitudinal momentum fractions of the jets.
The rapidities $y_{A,B}$ of the two tagged jets are related to the Sudakov parameters in the following way
\begin{equation}\label{eq:rapidities}
y_A=\frac{1}{2}\log{\left( \frac{x_{J_A}^2 s}{\vec{k}_A^2}\right)}\, \, , \, \,y_B=-\frac{1}{2}\log{\left( \frac{x_{J_B}^2 s}{\vec{k}_B^2}\right)}\, ,
\end{equation}
while the rapidity difference $Y$ is
\begin{equation}\label{eq:rapidities2}
Y\equiv y_A-y_B=\log{\frac{x_{J_A}x_{J_B}s}{\norm{\vec{k}_A}\norm{\vec{k}_B}}}\,.
\end{equation}

MN jets is a semi-hard process, in order to study it  properly we need to combine both collinear factorization and BFKL dynamics. Initially the two partons, before their hard partonic interaction which is described by BFKL, are following the standard DGLAP evolution~\cite{Gribov:1972ri,Altarelli:1977zs,Dokshitzer:1977sg}. 
In collinear factorization, the leading twist approximation allows us to write the cross section as a convolution of the parton distribution functions (PDFs) $f_i \left(x,\mu_F \right)$ and the partonic cross section $\hat{\sigma}$
\begin{equation}\label{eq:collfactorization}
\frac{\de{}{\sigma} \left( s \right)}{\de{}{y_A}\de{}{y_B}\de{2}{\vec{k}_A} \de{2}{\vec{k}_B}} = \sum_{i,j} \int_{0}^{1} f_i \left(x_A,\mu_F \right) f_j \left(x_B,\mu_F \right) \frac{\de{}{\hat{\sigma}_{ij}} \left(x_A x_B s, \mu_F \right)}{\de{}{y_A}\de{}{y_B}\de{2}{\vec{k}_A} \de{2}{\vec{k}_B}}
\end{equation}
where the indices $i,j$ specify the parton type $\left( i,j=q,\bar{q},g\right)$,  $\mu_F$ is the factorization scale
and
$x_{A,B}$ 
represent the longitudinal momentum fractions of the partons. Note that these are different from the jet momentum fractions and that $x_{J_{A,B}} \leq x_{A,B}$.

As mentioned earlier, within the BFKL framework, we can perform the resummation either at LLA or NLLA accuracy, however for both cases there is one additional important fact to keep in mind, namely, in the high-energy limit we are considering, the partonic cross-section itself also factorizes into a convolution of process-dependent jet vertices $V$ and a universal part which is accounted for by the gluon Green's function $\varphi$
\begin{multline}\label{eq:highenergyfactorization}
\frac{\de{}{\hat{\sigma}_{ij}} \left(x_A x_B s, \mu_F \right)}{\de{}{y_A}\de{}{y_B}\de{2}{\vec{k}_A} \de{2}{\vec{k}_B}}=
\frac{x_{J_A}x_{J_B}}{\left( 2\pi \right)^2} 
\int \frac{\de2{\vec{q}_A}}{\vec{q}_{A}^{\; 2}} V_i\left(\vec{q}_A,x_A,s_0,\vec{k}_A,x_{J_A}, \mu_F, \mu_R \right) 
\\ 
\int \frac{\de2{\vec{q}_B}}{\vec{q}_B^{\;2}} V_j\left(-\vec{q}_B,x_B,s_0,\vec{k}_B,x_{J_B}, \mu_F, \mu_R  \right) 
 \int_C \frac{\de{}{{\omega}}}{2\pi i} \left(\frac{x_A x_B s}{s_0} \right)^\omega \varphi_\omega\left( \vec{q}_A,\vec{q}_B\right)\,.
\end{multline}

Each jet vertex $V_{i (j)}$ describes the transition from the parton with longitudinal fraction $x_{A(B)}$ to the jet $A$($B$) after exchanging a $t$-channel Reggeized gluon with momentum $\vec{q}_{A(B)}$. The jet vertices depend on the factorization scale $\mu_F$, the renormalisation scale $\mu_R$,
an arbitrary energy scale $s_0$ that is introduced when taking the inverse Mellin transform in the calculation of the gluon Green's function and on the jet algorithm definition, whereas they have no $s$ dependence. It is important to note that the total cross section does not depend on $s_0$ within NLLA accuracy although $s_0$ affects higher order terms. 

The integration contour $C$ is a vertical line such that all poles in $\omega$ are to the left of the contour and the gluon Green's function satisfies the  forward BFKL equation
\begin{equation}\label{eq:bfkleqomega}
\omega \varphi_\omega\left( \vec{q}_A,\vec{q}_B\right) = \delta^2\left( \vec{q}_A-\vec{q}_B\right)+\int \de{2}{\vec{q}} K\left(\vec{q}_A,\vec{q} \right) \varphi_\omega \left( \vec{q},\vec{q}_B\right)\,.
\end{equation}

We will work with a combination of the jet vertex (impact factor)\footnote{Hereafter, the terms ``impact factor'' and ``jet vertex'' will be used interchangeably.}
 together with its respective contribution from of the PDFs, that is, with the inclusive impact factor:
\begin{equation}\label{eq:impactfactor}
\Phi\left(\vec{q},\vec{k},x_J, \omega, s_0,\mu_F, \mu_R \right) \equiv \sum_i \int_0^1 \de{}{x} f_{i}\left(x,\mu_F \right) V_i\left(\vec{q},x,s_0,\vec{k},x_{J}, \mu_F, \mu_R \right) \left(\frac{x}{x_J} \right)^\omega\, ,
\end{equation}
where the last factor $\left(\frac{x}{x_J} \right)^\omega$ is not conventional. It has been taken from the gluon Green's function to the impact factor to let us perform the integration over the parton momentum fraction $x$ before integrating over the rest of the variables, see~\cite{Caporale:2012ih}. This change does not affect the leading order (LO) part of the impact factor, since at this order we have $x=x_J$.

After using the inclusive impact factors $\Phi$, we write the differential cross section as
\begin{multline}\label{eq:totfactorization}
\frac{\de{}{\sigma} \left( s \right)}{\de{}{y_A}\de{}{y_B}\de{2}{\vec{k}_A} \de{2}{\vec{k}_B}} = 
\frac{x_{J_A}x_{J_B}}{\left( 2\pi \right)^2}  \int_C \frac{\de{}{{\omega}}}{2\pi i}
\int \frac{\de2{\vec{q}_A}}{\vec{q}_{A}^{\; 2}} \Phi\left(\vec{q}_A,\vec{k}_A,x_{J_A}, \omega, s_0,\mu_F, \mu_R \right)
\\ 
\int \frac{\de2{\vec{q}_B}}{\vec{q}_B^{\;2}}\Phi\left(-\vec{q}_B,\vec{k}_B,x_{J_B}, \omega, s_0,\mu_F, \mu_R \right)
e^{\omega\left( Y-Y_0 \right)} \varphi_\omega\left( \vec{q}_A,\vec{q}_B\right)\, ,
\end{multline}
where the relation between $Y_0$ and $s_0$ is given by $Y_0=\log{\frac{s_0}{\norm{\vec{k}_A}\norm{\vec{k}_B}}}$.

\subsection{The gluon Green's function at NLLA}
\label{sec:GGFNLLA}
In order to study the gluon Green's function at NLLA it is convenient to use the Dirac's braket notation as in~\cite{Caporale:2007vs} and to introduce eigenfunctions of the integral kernel, allowing us to solve the BFKL equation in a simple way. The explicit normalization and relationships of the eigenfunctions are found in Section \ref{ssec:normalization}. In operator notation, Eq.~(\ref{eq:bfkleqomega}) reads
\begin{equation}\label{eq:bfkleqoperator}
\omega \varphi_\omega = \ido + \calk \, \varphi_w
\end{equation}
and the total cross section (\ref{eq:totfactorization}) can be written as
\begin{equation} \label{eq:crosssectionoperators}
\frac{\de{}{\sigma} \left( s \right)}{\de{}{y_A}\de{}{y_B}\de{2}{\vec{k}_A} \de{2}{\vec{k}_B}} = 
\frac{x_{J_A}x_{J_B}}{\left( 2\pi \right)^2}  \int_C \frac{\de{}{{\omega}}}{2\pi i}
e^{\omega\left(Y-Y_0 \right)} \matrixel{\Phi(J_A)}{\frac{\ido}{\omega\ido-\calk}}{\Phi(J_B)}\, .
\end{equation}
Since the BFKL kernel is known to NLLA accuracy, we can write
\begin{equation}\label{eq:kernelexp}
\calk = \bam \calk_0 +\bam^2 \calk_1 + \calo\left(\bam^3 \right)\, ,
\end{equation}
where 
\begin{equation}\label{eq:alfas}
\bam=\frac{N_C \alpha_S(\mu_R^2)}{\pi}
\end{equation}
is the renormalized strong coupling constant evaluated at the renormalization scale $\mu_R$, while $N_C$ is the number of colors in QCD. $\calk_0$ and $\calk_1$ are the leading order (LO)~\cite{Balitsky:1978ic,Kuraev:1977fs,Kuraev:1976ge,Lipatov:1976zz,Fadin:1975cb} and next-to-leading order (NLO)~\cite{Fadin:1998py,Ciafaloni:1998gs} contributions to the BFKL kernel, which resum the leading and next-to-leading logarithms respectively.

At LLA the kernel enjoys conformal invariance and the eigenvectors can be found to be $\ket{n,\nu}$ (see Appendix \ref{ssec:normalization}) manifesting conformal symmetry~\cite{Lipatov:1985uk}. At NLLA this basis is no longer diagonal due to the breaking of conformal invariance by the running of the strong coupling and its action on the LLA eigenvectors is given by~\cite{Kotikov:2000pm}
\begin{equation}\label{LLAaction}
\matrixel{\vec{q}}{\calk_0}{n,\nu}=\chi_0(n,\nu)\braket{\vec{q}}{n,\nu},
\end{equation}
\begin{equation}\label{NLLAaction}
\matrixel{\vec{q}}{\calk_1}{n,\nu}=\chi_1(n,\nu)\braket{\vec{q}}{n,\nu}+\beta_0
 \left(\frac{i}{2} \frac{\del \chi_0(n,\nu)}{\del \nu} -\chi_0(n,\nu) \log{\frac{\vec{q}^{\;2}}{\mu_R^2}}  \right) \braket{\vec{q}}{n,\nu}\, ,
\end{equation}
where $\beta_0=\frac{11 N_C-2 N_F}{12 N_C}$ is the one-loop beta function of the strong coupling and $N_F$ is the number of active quark flavors. Note that the second term in the r.h.s. of Eq.~(\ref{LLAaction})  is proportional to $\beta_0$ and it is either purely imaginary or $\vec{q}^{\;2}$ dependent.

The LO kernel eigenvalue $\chi_0(n,\nu)$ is given by
\begin{equation}\label{eq:LOeigenvalue}
\chi_0(n,\nu)=2\psi(1)- \psi\left(\frac{1+\abs{n}}{2}+i \nu \right)-\psi\left(\frac{1+\abs{n}}{2}-i \nu \right)\, ,
\end{equation}
with $\psi(z)=\frac{\de{}{}}{\de{}{z}} \log{\Gamma(z)}$ the digamma function. 

The remaining term $\chi_1(n,\nu)$ is given by~\cite{Kotikov:2000pm}
\begin{multline}\label{eq:NLOeigenvalue}
\chi_1(n,\nu)=\gamma_K^{(2)} \chi_0(n,\nu) + \frac{3}{2} \zeta(3)-\frac{\beta_2}{2}\chi_0^2(n,\nu)+\frac{1}{4}\chi''_0(n,\nu)
-\frac{1}{2}\left( \varPhi(\abs{n},\nu)+\varPhi(\abs{n},-\nu)    \right)+
\\
+\frac{\pi^2 \sinh{\pi \nu}}{8 \nu \cosh^2{\pi \nu}} \left\lbrace   
-\delta_{n 0} \left[ 3+\left(1+\frac{N_F}{N_C^3} \right) \frac{11+12\nu^2}{16\left(1+\nu^2 \right)}   \right] 
+\delta_{\abs{n} 2} \left(1+\frac{N_F}{N_C^3} \right) \frac{1+4\nu^2}{32\left(1+\nu^2 \right)}
   \right\rbrace.
\end{multline}
\text{In Eq.~(\ref{eq:NLOeigenvalue}) $\varPhi(n,\nu)$ is defined as}
\begin{multline}\label{eq:Phidef}
\varPhi(n,\nu)=\sum_{k=0}^\infty \frac{(-1)^{k+1}}{k+i\nu+\frac{1+n}{2}} \bigg\lbrace
\psi'(k+n+1)-\psi'(k+1)+ 
\\
\left.  +(-1)^{k+1}\left( \beta'(k+n+1)+\beta'(k+1) \right) + \frac{\psi(k+1)-\psi(k+n+1)}{k+i\nu+\frac{1+n}{2}}
\right\rbrace \, ,
\end{multline}
where
\begin{equation}\label{eq:defbetaf}
\beta(z)=\frac{1}{2}\left( \psi\left(\frac{1+z}{2}\right)- \psi\left(\frac{z}{2}\right)   \right)\, ,
\end{equation}
while the accents on $\chi_0(n,\nu)$ indicate derivatives with respect to $\nu$. The two-loop QCD cusp anomalous dimension in the dimensional reduction scheme is
\begin{equation}\label{eq:cuspdef}
\gamma_K^{(2)}=\frac{1}{3}\left(5 \beta_2+1\right)-\frac{\zeta(2)}{2}=\frac{1}{4}\left(\frac{67}{9}-\frac{10 N_F}{9 N_C} -2 \zeta(2) \right)\, .
\end{equation}

In~\cite{Chirilli:2013kca,Chirilli:2014dcb}, Chirilli and Kovchegov built the eigenvectors of the NLLA kernel perturbatively, expanding around the LLA (conformal) ones. Their detailed properties are given in Appendix \ref{ssec:normalization}. Here we will only note that they satisfy
\begin{equation}\label{NLOeigenaction}
\calk \ket{H_{n,\nu}}= \left(\bam \chi_0(n,\nu)+ \bam^2 \chi_1(n,\nu)\right)\ket{H_{n,\nu}} \equiv \chi(n,\nu) \ket{H_{n,\nu}}\,\,.
\end{equation}

The remaining necessary ingredient to compute the cross section is the impact factor which was calculated at NLO in~\cite{Bartels:2001ge,Bartels:2002yj} and was later confirmed in~\cite{Caporale:2011cc}. We will be using the small-cone approximation (SCA), where the jet cone aperture $R$ in the rapidity-azimuthal angle plane is considered small, neglecting powers in $R$. The impact factor was calculated in Ref.~\cite{Ivanov:2012ms} directly in $\ket{n,\nu}$ space where the result can be expressed in a simple analytical form.  A comparison between different jet algorithms (the Furman algorithm~\cite{Furman:1981kf}, the $k_T$  algorithm~\cite{Catani:1993hr} and the cone algorithm~\cite{Ellis:1989vm}) can be found in~\cite{Colferai:2015zfa}.

The expressions for the LO and NLO impact factor can be directly extracted from Ref.~\cite{Ivanov:2012ms} after taking into account some slight changes in normalization and notation. They respectively read
\begin{equation}\label{eq:LOimpactfactor}
\braket{\Phi^{(LO)}\left( J \right)}{n,\nu} = \frac{\caln}{\vec{k}_J^2} f^{\star}(x_J)\frac{1}{\pi \sqrt{2}}\left(\vec{k}^{2} \right)^{i \nu-\frac{1}{2}} e^{i n \theta_J}=\frac{\caln}{\vec{k}_J^2} f^{\star}(x_J) \braket{\vec{k}}{n,\nu}
\end{equation}
and
\begin{equation}\label{eq:NLOimpactfactor}
\braket{\Phi^{(NLO)}\left( J \right)}{n,\nu} =\frac{\caln}{\vec{k}_J^2} f^{\star}(x_J) \braket{\vec{k}}{n,\nu} \left(1+\bam \phi_1 \left( n,\nu,\omega,J \right)  \right)\, .
\end{equation}
The normalization factor is $\caln=2 \pi^2 \bam \sqrt{\frac{2 C_F}{N_C^3}}$ and the NLO correction to the impact factor, $\phi_1 \left( n,\nu,J \right)$, can be found in 
Appendix~\ref{ssec:nloimpactfacor}. We have also defined the effective PDF~\cite{Combridge:1983jn}
\begin{equation}\label{eq:effectivepdf}
f^{\star}(x_J)\equiv \frac{N_C}{C_F} f_g (x_J)+\sum_{i=q,\bar{q}} f_i (x_J)\, .
\end{equation}

The LO and NLO impact factor in the $\ket{H_{n,\nu}}$ basis can be easily computed, resulting in
\begin{equation}\label{eq:LOimpactfactorH}
\braket{\Phi^{(LO)}\left( J \right)}{H_{n,\nu}} = \frac{\caln}{\vec{k}_J^2} f^{\star}(x_J) \braket{\vec{k}}{H_{n,\nu}}
\end{equation}
and
\begin{equation}\label{eq:NLOimpactfactorH}
\braket{\Phi^{(NLO)}\left( J \right)}{H_{n,\nu}} =\frac{\caln}{\vec{k}_J^2} f^{\star}(x_J) \braket{\vec{k}}{H_{n,\nu}} \left(1+\bam \phi_1 \left( n,\nu,\omega,J \right)  \right)
\end{equation}
respectively. 

The cross section after using the NLLA kernel and the LO impact factors reads
\begin{multline}\label{eq:crosssectionoperatorLO}
\frac{\de{}{\sigma^{(LO,NLLA)}} \left( s \right)}{\de{}{y_A}\de{}{y_B}\de{2}{\vec{k}_A} \de{2}{\vec{k}_B}} = 
\frac{x_{J_A}x_{J_B}}{ 2\pi } \sum_n \int \frac{\de{}{\nu}}{2 \pi}
e^{\chi(n,\nu)\left(Y-Y_0 \right)} \braket{\Phi^{(LO)}(J_A)}{H_{n,\nu}}\braket{H_{n,\nu}}{\Phi^{(LO)}(J_B)}=
\\
=\frac{\caln^2 x_{J_A}x_{J_B}}{ 2\pi \vec{k}_A^2 \vec{k}_B^2} f^{\star}(x_{J_A}) f^{\star}(x_{J_B}) 
\sum_n \int \frac{\de{}{\nu}}{2 \pi}
e^{\chi(n,\nu)\left(Y-Y_0 \right)} \braket{\vec{k}_A}{H_{n,\nu}}\braket{H_{n,\nu}}{-\vec{k}_B}  =
\\
=\frac{4 \pi^2 C_F \bam^2 }{ N_C^3 \norm{\vec{k}_A}^3 \norm{\vec{k}_B}^3} x_{J_A} f^{\star}(x_{J_A}) x_{J_B} f^{\star}(x_{J_B}) 
\sum_n \frac{e^{i n \theta}}{2 \pi} \int \frac{\de{}{\nu}}{2 \pi}
e^{\tilde{\chi}(n,\nu)\left(Y-Y_0 \right)}\left( \frac{\vec{k}_A^2}{\vec{k}_B^2}\right)^{i \nu} \,,
\end{multline}
where the azimuthal angle difference\footnote{When $\theta=0$ the two jets are back to back in transverse space.} $\theta$  is given by $\theta\equiv \theta_A-\theta_B-\pi$ and we have also inserted the term $\ket{H_{n,\nu}} \bra{H_{n,\nu}}$ . 
  Following Ref.~\cite{Chirilli:2013kca}, we have placed the terms originated by the NLO eigenfunctions in the exponential, thus modifying the eigenvalue to
\begin{equation}
\tilde{\chi}(n,\nu)   \equiv  \bam \chi_0(n,\nu) \left(1-\bam \beta_2 \log{\frac{\norm{\vec{k}_A}\norm{\vec{k}_B}}{\mu_R^2}} \right)+ \bam^2 \chi_1(n,\nu)\,.
\end{equation}
If we neglect higher order terms beyond NLO, we can interpret the modification of the eigenfunctions as a change of the renormalization scale from $\mu_R$ to $\mu_N = \sqrt{\norm{\vec{k}_A} \norm{\vec{k}_B}}$ (natural scale) and therefore, we can use the LO eigenfunctions instead of the NLO ones after setting the renormalization scale in the kernel to be equal to the natural scale $\mu_N$.
Following the same steps for the case of the NLO impact factor, we finally obtain 
\begin{multline}\label{eq:crosssectionNLO}
\frac{\de{}{\sigma^{(NLO,NLLA)}} \left( s \right)}{\de{}{y_A}\de{}{y_B}\de{2}{\vec{k}_A} \de{2}{\vec{k}_B}} 
=\frac{4 \pi^2 C_F \bam^2 }{ N_C^3 \norm{\vec{k}_A}^3 \norm{\vec{k}_B}^3} x_{J_A} f^{\star}(x_{J_A}) x_{J_B} f^{\star}(x_{J_B}) 
\\
\sum_n \frac{e^{i n \theta}}{2 \pi} \int \frac{\de{}{\nu}}{2 \pi}
\left( \frac{\vec{k}_A^2}{\vec{k}_B^2}\right)^{i \nu}  \left(1+\bam \phi_1(n,\nu,\chi,J_A) \right) \left(1+\bam \bar{\phi}_1(n,\nu,\chi,J_B) \right)
e^{\tilde{\chi}(n,\nu)\left(Y-Y_0 \right)}\,.
\end{multline}

\subsection{Minijet radiation after imposing a rapidity veto between subsequent emissions}

Imposing the constraint that subsequent minijet emissions must have a rapidity difference greater than a fixed value $b$, the rapidity veto, leads to the following modification of the BFKL kernel at LLA~\cite{Schmidt:1999mz} which only affects terms beyond LLA accuracy: 
\begin{multline}\label{eq:GGFvetoLOexp}
\varphi\left(\vec{q}_A,\vec{q}_B,Y \right)=\int_\calc \frac{\de{}{\omega}}{2 \pi i} e^{\omega (Y-Y_0-b)} \matrixel{\vec{q}_A}{\frac{\ido}{\omega \ido-e^{-\omega b}\calk}}{\vec{q}_B}=
\\
=\frac{2}{ \norm{\vec{q}_A} \norm{\vec{q}_B}} \sum_n \frac{ e^{i n (\theta_{q_{A}}-\theta_{q_{B}})}}{2 \pi}\int \frac{\de{}{\nu}}{2 \pi }  \left( \frac{\vec{k}_A^2}{\vec{k}_B^2}\right)^{i \nu}   \frac{e^{\tilde{\omega} (Y-Y_0-b)} }{1+b \tilde{\omega}}\,,
\end{multline}
where $\tilde{\omega}$ is by definition the solution to the following transcendental equation
\begin{equation}\label{eq:transcendeltaLO}
\tilde{\omega}=e^{-b \tilde{\omega}} \bam \chi_0 (n,\nu) \leftrightarrow \tilde{\omega}=\frac{\calw\left(b \, \bam \chi_0(n,\nu)\right)}{b}\,\,,
\end{equation}
where $\calw$ is Lambert's W function. As pointed out in~\cite{Forshaw:1999xm}, the solution of (\ref{eq:transcendeltaLO}) develops unphysical branch points at $\pm \nu_0$ that satisfy
\begin{equation}
\bam \chi_0(n,\nu_0)=-\frac{e^{-1}}{b}
\end{equation}
because we have summed over an arbitrarily large number of gluon emissions, something that is inconsistent with the rapidity veto constraint. We can expand the gluon Green's function as a power series in the kernel and truncate the sum to limit the number of emissions 
\begin{equation}\label{eq:GGFvetoLOser}
  \frac{e^{\tilde{\omega} (Y-Y_0-b)} }{1+b \tilde{\omega}} =  \sum_k \frac{ \left( Y-Y_0-(k+1)b \right)^k \left(\bam \chi_0(n,\nu) \right)^k }{k!}\,.
\end{equation}

The power series converges only asymptotically~\cite{Schmidt:1999mz} and the best approximation is obtained by the truncated series at the largest value of $k$ that satisfies $\left( Y-Y_0-(k+1)b \right) > 0$. This is in accordance with the physical intuition that only a fixed number of emissions should be allowed if the rapidity constrained is to be respected. In our numerical computations and the results we will present in the next section, we have used the truncated expansion.

The modification of the BFKL equation after imposing a rapidity veto at NLLA was also carried out by Schmidt in~\cite{Schmidt:1999mz}. The $b$ dependence can be set requiring that the total cross section only depends on $b$ through NNLLA terms, \ie, the corrections to the cross section of order $ (\bam Y)^n$ and $\bam (\bam Y)^n$ should be independent on the veto. In that way, the influence of the veto at NLLA is reduced considerably compared to the LLA case. The veto dependence is found after performing the following modifications to the kernel and to the impact factors
\begin{equation}\label{eq:modkernelVeto}
\calk \rightarrow \calk_b = \calk + b  \calk_0 \calk_0 +\calo(\bam^3)
\end{equation}
\begin{equation}\label{eq:modimpactVeto}
\ket{\Phi(J)}\rightarrow \ket{\Phi_b(J)} =\ket{\Phi(J)} + b \calk_0  \ket{\Phi^{(LO)}(J)}+\calo(\bam^3)
\end{equation}
 and the total cross section can be written as
 \begin{equation} \label{eq:crosssectionVeto}
\frac{\de{}{\sigma} \left( s \right)}{\de{}{y_A}\de{}{y_B}\de{2}{\vec{k}_A} \de{2}{\vec{k}_B}} = 
\frac{x_{J_A}x_{J_B}}{\left( 2\pi \right)^2}  \int_C \frac{\de{}{{\omega}}}{2\pi i}
e^{\omega\left(Y-Y_0-b \right)} \matrixel{\Phi_b(J_A)}{\frac{\ido}{\omega\ido-e^{-b \omega}\calk_b}}{\Phi_b(J_B)}\,.
\end{equation}

Using the completeness relation for $\ket{H_{n,\nu}}$, we can solve the equation as previously resulting in
\begin{multline}\label{eq:crosssectionoperatorVetoNLO}
\frac{\de{}{\sigma^{(NLO,NLLA)}} \left( s \right)}{\de{}{y_A}\de{}{y_B}\de{2}{\vec{k}_A} \de{2}{\vec{k}_B}} 
=\frac{4 \pi^2 C_F \bam^2 }{ N_C^3 \norm{\vec{k}_A}^3 \norm{\vec{k}_B}^3} x_{J_A} f^{\star}(x_{J_A}) x_{J_B} f^{\star}(x_{J_B}) 
\sum_n \frac{e^{i n \theta}}{2 \pi} \int \frac{\de{}{\nu}}{2 \pi} 
\left( \frac{\vec{k}_A^2}{\vec{k}_B^2}\right)^{i \nu} 
\\
\frac{e^{\tilde{\omega}' (Y-Y_0-b)} }{1+b \tilde{\omega}'}
 \left(1+\bam \phi_1(n,\nu,\chi,J_A) + \bam b \chi_0(n,\nu )\right)
  \left(1+\bam \bar{\phi}_1(n,\nu,\chi,J_B)+ \bam b \chi_0(n,\nu) \right)\,,
\end{multline}
where now
\begin{equation}\label{eq:transcendeltaNLO}
\tilde{\omega}'=e^{-b \tilde{\omega}'} \tilde{\chi}_b(n,\nu) \leftrightarrow \tilde{\omega}'=\frac{\calw\left(b \tilde{\chi}_b(n,\nu) \right)}{b}
\end{equation}
and
\begin{equation}
\tilde{\chi}_b(n,\nu)   =  \bam \chi_0(n,\nu) \left(1+\bam b\chi_0(n,\nu) -\bam \beta_2 \log{\frac{\norm{\vec{k}_A}\norm{\vec{k}_B}}{\mu_R^2}} \right)+ \bam^2 \chi_1(n,\nu)\,.
\end{equation}
Repeating the procedure we followed in the LLA case, we expand to get the asymptotic sum
\begin{equation}\label{eq:GGFvetoNLOser}
  \frac{e^{\tilde{\omega}' (Y-Y_0-b)} }{1+b \tilde{\omega}'} =  \sum_k \frac{ \left( Y-Y_0-(k+1)b \right)^k \left(\tilde{\chi}_b(n,\nu) \right)^k }{k!}\,.
\end{equation}
The final expression we use for our numerical computations  is (\ref{eq:crosssectionoperatorVetoNLO}) combined with (\ref{eq:GGFvetoNLOser}) and integrated over the appropriate phase space.

\subsection{Azimuthal decorrelation coefficients}

The original proposal by Mueller and Navelet~\cite{Mueller:1986ey} was to study the dependence of the cross section on the increasing rapidity difference $Y$, however, it proved more advantageous to study the azimuthal decorrelation of the two jets as was proposed in~\cite{DelDuca:1993mn,Stirling:1994he}. The total cross section receives large corrections after including the NLLA terms to the gluon Green's function largely increasing the theoretical uncertainty. This can, to a major degree, be avoided by studying ratios of azimuthal decorrelation coefficients~\cite{Vera:2007kn} which essentially means removing the contribution with conformal spin $n=0$. This leads to theoretical computations with much better perturbative stability. In this work, since we are interested in investigating the fundamental character of the rapidity veto we will include observables that depend on the zeroth conformal spin and study the influence they show on the veto. 

 The Fourier expansion in the azimuthal angle difference $\theta$ of
the differential cross section is 
 \begin{equation} \label{eq:crosssectioncosines}
\frac{\de{}{\sigma} (s)}{\de{}{y_A}\de{}{y_B}\de{}{\norm{\vec{k}_A}} \de{}{\norm{\vec{k}_B}}\de{}{\theta_A} \de{}{\theta_B}} = 
\frac{1}{\left(2\pi \right)^2} \left(\calc_0 + \sum_{n=1}^{\infty} 2 \cos{(n \theta )} \;  \calc_n \right)\,,
\end{equation}
and therefore the azimuthal decorrelation is directly related to the coefficients $\frac{\calc_n}{\calc_0}=\left\langle \cos{(n \theta )} \right\rangle$.

The physical intuition about the decorrelation coefficients is the following. If there are only the two tagged jets in the final state, they will be totally correlated due to momentum conservation, making all $\frac{\calc_n}{\calc_0}=1$. Due to the extra radiation, the distribution in $\theta$ is not a delta, and its moments are fully characterized by the coefficients, providing information about the importance of the minijet radiation that we have not tagged. In the BFKL approach, an increase in $Y$ will lead to an increase in the amount of radiation\footnote{This argument is not valid at the boundary of the phase space, when there is no more available energy to produce extra radiation.}, generating more decorrelation on the tagged jets.

\section{Behaviour of the azimuthal coefficients after imposing the veto and comparison to experimental data }
\subsection{Kinematics and the specifics of the numerical analysis }

In order to compare against experimental data, a phase space integration over $\vec{k}_A$ and $\vec{k}_B$
is needed. The measurement of Mueller-Navelet azimuthal decorrelation has been performed by the CMS collaboration in~\cite{CMS:2013eda,Khachatryan:2016udy}, and by ATLAS in~\cite{Aad:2011jz}. We will use the kinematical cuts of the CMS 
analysis, in particular for our numerical study we use
\begin{equation}\label{eq:phasespace}
35 \text{ GeV} \leq \norm{\vec{k}_A}, \norm{\vec{k}_B} \leq 60 \textrm{ GeV} \; \; , \; \;  \abs{\norm{\vec{k}_A}-\norm{\vec{k}_B}} \geq 2 \text{ GeV} \; \; , \; \; 0 \leq y_A, \abs{y_B} \leq 4.7\,,
\end{equation}
whereas the initial observables we are computing are the following
\begin{equation}\label{eq:coeffphasespace}
C_n (Y,b)= \int \de{2}{\vec{k}_A} \de{2}{\vec{k}_B} \de{}{y_A}\de{}{y_B} \delta(y_A-y_B-Y) \cos{(n \theta)} 
\frac{\de{}{\sigma}}{\de{}{y_A}\de{}{y_B}\de{2}{\vec{k}_A} \de{2}{\vec{k}_B}}
\end{equation}
in order to compute the final ratios
\begin{equation}\label{eq:ratiosdef}
R^n_m (Y,b) \equiv \frac{C_n (Y,b)}{C_m (Y,b)}\,.
\end{equation}

Some remarks are in order at this point:

\begin{itemize}
\item In the CMS analysis~\cite{CMS:2013eda,Khachatryan:2016udy}, in the relevant plots, what is shown is not $R^n_m (Y,0)$ but rather $R^n_m (Y,0)$ integrated over a bin in the rapidity difference centered at $Y$. The latter would be preferable in the theoretical analysis of this study for a one to one comparison, however, such a computation requires one further integration and the accompanying computational cost. We have decided to avoid this extra integration since the purpose of the current investigation is to understand the influence of the veto on the azimuthal decorrelations rather than performing a proper fitting to the experimental points. Moreover, we should note that since the total cross section decreases quickly with the rapidity difference, the result from integrating over any given rapidity bin will be 
biased toward the value at the smaller limit of the rapidity bin. 

\item There is no upper transverse momenta cutoff in the experimental selection. This is done for numerical reasons, but the dependence of the observable on this parameter is negligible, since the cross section is rapidly decreasing with increasing transverse momenta as was demonstrated in~\cite{Ducloue:2013wmi}.

\item The phase space region considered in our analysis has a lower cutoff in the difference of the transverse momenta $| \norm{\vec{k}_A}-\norm{\vec{k}_B}|$. This reduces the influence of collinear contamination effects on the observables. Actually, it has been suggested~\cite{Caporale:2014gpa} that eliminating the ``back to back''  region will enhance the BFKL effects with respect to fixed order calculations. 

\item The jet algorithm used in the experimental analysis in~\cite{CMS:2013eda,Khachatryan:2016udy} was the anti-$k_T$ clustering algorithm with a radius value $R=0.5$, while here we have used the Furman algorithm with $R=0.5$ as we have already discussed in Section \ref{sec:GGFNLLA} (see discussion in Ref.~\cite{Colferai:2015zfa}).

\item There is a mismatch in the event selection procedure from the experimental and the theoretical side as was originally pointed out in~\cite{Caporale:2014gpa}. To make that clear, let us suppose that we have three jets satisfying the transverse momenta conditions, two of them in the very forward region with rapidities $y_1$,$y_2$ ($y_1>y_2$) while the third one is very backward with a rapidity $y_3$. In the CMS study this would be counted as one MN event with tagged jets the ones with rapidities $y_1$ and $y_3$. However, another MN configuration is possible here formed by the jets with rapidities $y_2$ and $y_3$. The effect from this discrepancy  has been computed in~\cite{Colferai:2017bog}, and it is below the $4\%$ level with a peak at rapidity difference $Y\simeq4$.
\end{itemize}
 
Finally, in order to compare against the experimental values, a center-of-mass energy of $\sqrt{s}=7 \text{ TeV}$ was used while the rapidity scale $Y_0$ was set to zero. The NLO MSTW 2008 PDF\footnote{Changing 
from  MSTW08 to MMHT14 has a very small impact on our numerical results, typically, less than 1\permil.} 
 sets~\cite{Martin:2009iq} were used, while for the strong coupling $\alpha_s$ we chose a two-loop running coupling setup with $\alpha_s (M_Z) = 0.11707$ setting the number of active flavors to $N_F=5$. The renormalization scale was chosen to be the natural one, $\mu_R=\mu_N=\sqrt{\norm{\vec{k}_A}\norm{\vec{k}_B}}$. The factorization scales have been set to $\mu_{F_{A,B}}=\norm{\vec{k}_{A,B}}$ for each impact factor respectively. The influence of using such a choice instead of $\mu_{F_{A,B}}=\mu_R$ was investigated in~\cite{Caporale:2014gpa}.  Finally, the multivariable integrations have been performed using the numerical integration packages in \cod{MATHEMATICA}.

\subsection{Results}
In this Section we present our results in Figs.~$1-8$.  In each figure, we plot the dependence of some of the ratios defined in \ref{eq:ratiosdef} on the veto $b$ for a fixed value of the rapidity difference $Y$.  In particular, we show results for $R^1_0 (Y,b) $, $R^2_0 (Y,b)$, $R^3_0 (Y,b)$, $R^2_1 (Y,b)$ and $R^3_2 (Y,b)$. In each plot we include the experimental value along with its systematic uncertainty (reddish fixed-width band) for the rapidity bin centered at the given rapidity point. The experimental values have been obtained from the 2013~\cite{CMS:2013eda} analysis, since the 2016 ones~\cite{Khachatryan:2016udy} are not yet publicly available. This has special impact on the $Y=5.75$ data, since this point shows an increase on the value of $R^3_0 (Y,b)$ and $R^3_2 (Y,b)$ that is not shown in the 2016 analysis. The theoretical uncertainty from the variation of the renormalization scale\footnote{The numerical uncertainty of our computations  is negligible compared to the scale variation uncertainty.} is represented by a bluish band the limits of which are obtained for $\mu_R = \mu^{\text{centr}}_R/2$ and 
$\mu_R =2 \mu^{\text{centr}}_R$, where $\mu^{\text{centr}}_R$ is the renormalisation scale used for the computation of the central blue dashed line within the bluish uncertainty band.

\plotall{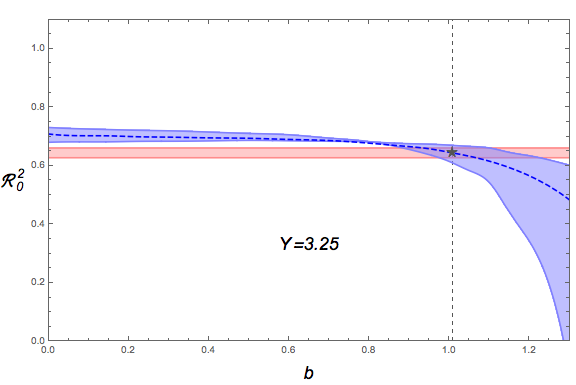}{3.25}
\plotall{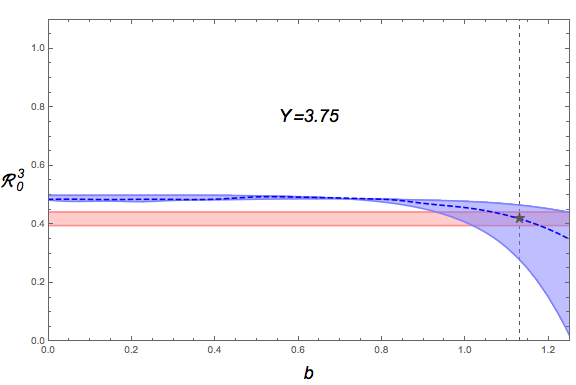}{3.75}
\plotall{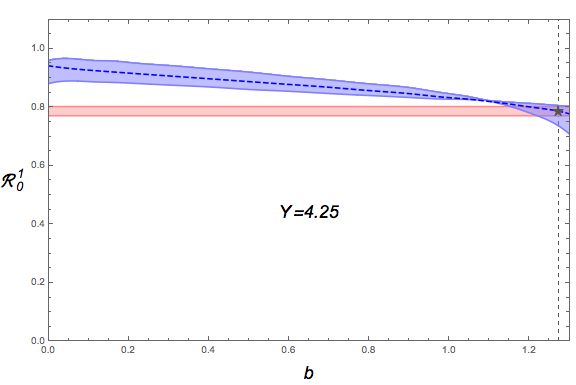}{4.25}
\plotall{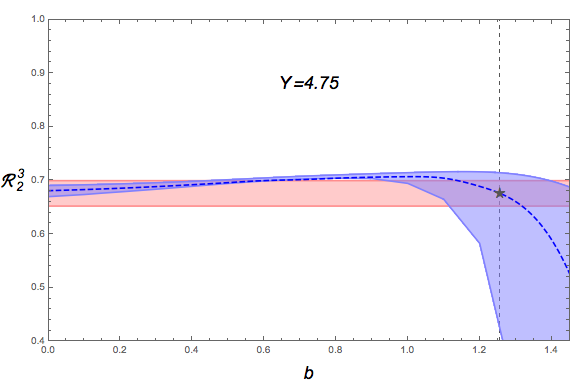}{4.75}
\plotall{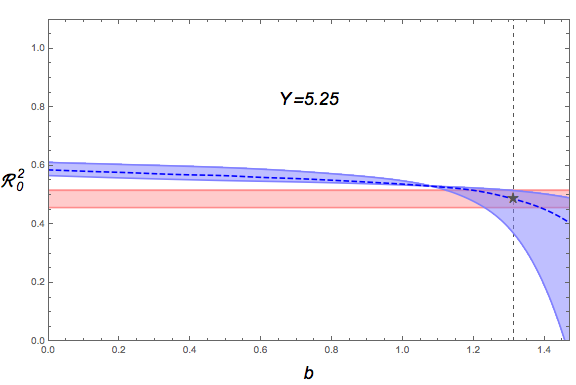}{5.25}
\plotall{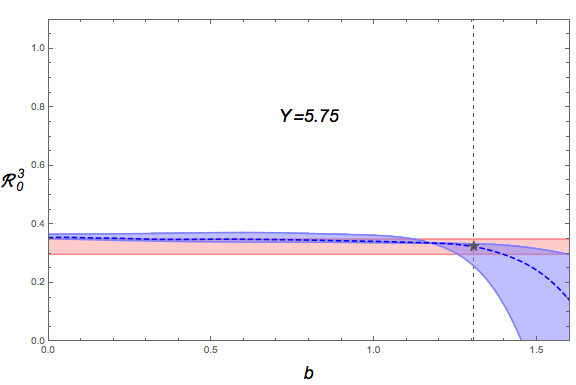}{5.75}
\plotall{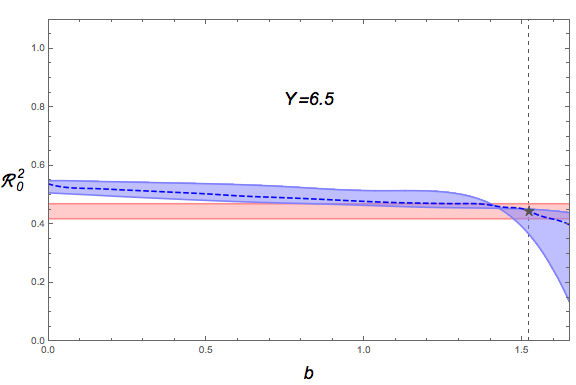}{6.5}

There are several features that are common to all plots. In the regions with larger veto values the coefficients become negative since they are more affected by  collinear contributions. This effect is more pronounced when we reduce the renormalization scale (since we approach the non-perturbative regime).  
Except for the large rapidity region ($Y=7.5$ and $8.7$) for which our expansion breaks down and for which we will not show any plots, there is always an interval of the veto where not only the ratios of correlation functions ($R^2_1 (Y,b)$ and $R^3_2 (Y,b)$) but also the correlation functions themselves ($R^1_0 (Y,b) $, $R^2_0 (Y,b)$, $R^3_0 (Y,b)$) are well described. 
Interestingly, this happens for values of the veto in the region where the asymptotic series we are using reaches its boundary of convergence ($b \gtrsim 1$). 
A natural explanation for the unfavourable behaviour of the veto approach at large $Y$ and for the present setup 
should be sought at the influence of the PDFs in that regions. More concretely, we are in a region where the PDFs tend very fast to zero and introduce extra energy-momentum conservation effects beyond those present in the BFKL Green's function. This is the reason why the idea of a simple constant rapidity veto is not viable at larger $Y$.  

\begin{figure}[h]
\begin{center}
\includegraphics[width=8.2cm]{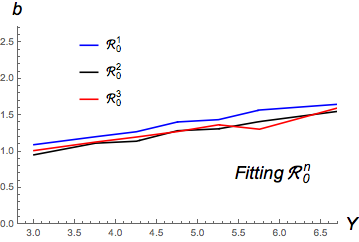} \includegraphics[width=8.2cm]{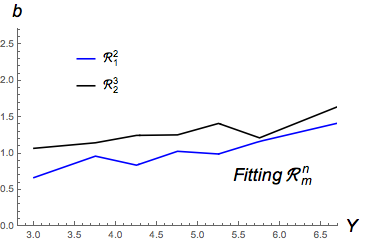} 
\end{center}
\caption{Dependence of the rapidity veto best-fit values on the rapidity $Y$ for the different $R^n_m$ functions.}
\label{fit}
\end{figure}

Fig.~\ref{fit} shows the values of the rapidity veto that fit best the experimental values for the correlation functions (left) and the ratios of correlation functions (right) in the rapidity range $3.25 \le Y \le 6.5$.
We find that the optimal value of the veto fitting the data slightly grows monotonically with the rapidity difference $Y$.

Returning to the questions we have listed in the introduction, it is safe to claim that is it possible to obtain a good theoretical description of the ratios ${\cal R}^m_n$ including the ones with either $m=0$ or $n=0$ without necessarily using the BLM scheme. Moreover, a rapidity veto allows in principle for a fit of the data --excluding the larger $Y$ ones as explained previously-- assuming that  $b \gtrsim 1$.   Regarding the question on whether the optimal $b$ values tell us how far from asymptotia we are at LHC energies, one can only make qualitative statements. Clearly, at LHC energies we are not in asymptotia, however, a value of $b$ close to unity suggests that pre-asymptotic BFKL effects are already present and possibly important.

\section{Conclusions}

There are different methods to stabilize the perturbative expansion in the Multi-Regge kinematics regime. Here we have explored the possibility of introducing a rough constant cut-off in the rapidity differences among emitted mini jets in the final state. By comparing to current LHC data we have found that it is possible to get a reasonable global description of many different azimuthal angle correlations in dijet cross sections with a rapidity veto $b \gtrsim 1$. This value is far from previous formal studies based on the asymptotic behavior of the gluon Green's function alone and it depends on the actual rapidity difference. The effect of introducing jet vertices and parton distribution functions at realistic energies such as those proved at the LHC is to drastically reduce the value of the veto from $b \gtrsim 2$ to half of this value. This encodes different types of information. It shows how far we are in this observable from the asymptotic region where power-suppressed with energy terms are negligible. It also indicates the size of the collinear regions of phase space which need to be taken into account in order to get a good description of the data. The success of renormalization schemes such as the BLM approach when applied to the class of observables here described can be also understood since they essentially absorb the effect of the veto using a redefinition of the position of the Landau pole which numerically generates similar values of the cross sections.

\section*{Acknowledgements}

This work was supported by the Spanish Research Agency (Agencia Estatal de Investigaci\'on) through the grant IFT Centro de Excelencia Severo Ochoa SEV-2016-0597. GC and ASV acknowledge support from the Spanish Government grants FPA2015-65480-P, FPA2016-78022-P. DGG is supported with a fellowship of the international programme `La Caixa-Severo Ochoa'. FGC acknowledges support from the Italian Foundation ``Angelo della Riccia''.

\begin{appendices}

\section{}
\subsection{Normalization of the eigenvectors}
\label{ssec:normalization}

\begin{equation}
\braket{\vec{q}_A}{\vec{q}_B}=\delta^2\left(\vec{q}_A-\vec{q}_B \right) \;\; , \;\; \ido=\int \de{2}{\vec{q}} \ket{\vec{q}} \bra{\vec{q}}
\end{equation}

\begin{equation}
\braket{n,\nu}{m,\mu}=\delta\left(\nu-\mu \right)\delta_{nm}  \;\; , \;\; \ido=\sum_{n}\int \de{}{\nu} \ket{n,\nu} \bra{n,\nu}
\end{equation}

\begin{equation}
\braket{H_{n,\nu}}{H_{m,\mu}}=\delta\left(\nu-\mu \right)\delta_{nm} + \calo\left(\bam^2 \right)  \;\; , \;\; \ido=\sum_{n}\int \de{}{\nu} \ket{H_{n,\nu}} \bra{H_{n,\nu}}+ \calo\left(\bam^2 \right)
\end{equation}

\begin{equation}
\braket{\vec{q}}{n,\nu}=\frac{1}{\pi \sqrt{2}} \left(\vec{q}^{\;2} \right)^{i \nu-\frac{1}{2}} e^{i n \theta}
\end{equation}

\begin{equation}
\braket{\vec{q}}{H_{n,\nu}}=\frac{1}{\pi \sqrt{2}} \left(\vec{q}^{\;2} \right)^{i \nu-\frac{1}{2}} e^{i n \theta} \left[ 
1+ \bam \beta_2 \log{\frac{\vec{q}^{\;2}}{\mu_{R}^2}} \left(   
A\left( n,\nu \right) \log{\frac{\vec{q}^{\;2}}{\mu_{R}^2}}+B\left( n,\nu \right) 
\right) \right]
\end{equation}
where $A\left( n,\nu \right)=\frac{i}{2} \frac{\chi_0 \left(n,\nu \right)}{ \chi'_0 \left(n,\nu \right)}$ and $B\left( n,\nu \right)=\frac{1}{2} \frac{\del}{\del \nu}\frac{\chi_0 \left(n,\nu \right)}{\chi'_0 \left(n,\nu \right)}$. The prime indicates a derivative with respect to $\nu$ and the integrals over $\nu$ must be regulated at $\nu=0$ treating the eigenfunctions as distributions, and applying the principal value prescription to the pole.

\subsection{NLO Impact factor}
\label{ssec:nloimpactfacor}

We have used the NLO impact factor calculated in \cite{Ivanov:2012ms}, being aware that their definition of $\gamma$, $\gamma_{I.P.}= -\frac{1}{2}+ i \nu $, is different from ours, $\gamma= \frac{1}{2}+ i \nu $. Also, some simplifications were done converting the hypergeometric functions into incomplete Beta functions. We remind that the effective PDF is defined as $
f^{\star}(x_J)\equiv \frac{N_C}{C_F} f_g (x_J)+\sum_{i=q,\bar{q}} f_i (x_J)$.

\begin{equation}
\begin{aligned}
\phi_{1} (n,\gamma,\omega,J) = \frac{f_q (x_J, \mu_F)}{2 C_A f^\star (x_J,\mu_F)} \left\lbrace
C_q + 2(C_F-C_A) \log{(1-x_J)}^2+C_F (x_J^2+2 x_J +4 \log{(1-x_J)})\log{\frac{k_J}{R \mu_F}}
 \right\rbrace + \\
+ \frac{f_g (x_J, \mu_F)}{2 C_F f^\star (x_J,\mu_F)}      \left\lbrace
C_g + 4 N_C \beta_2 \log{\frac{2 k_J}{R  \mu_F}}+2 C_A  \log{(1-x_J)} (2 \log{\frac{k_J}{R \mu_F}}+\chi_0(n,\gamma))
 \right\rbrace + \\
 + \frac{1}{4}\left\lbrace  
 \psi'(\frac{n}{2}+1-\gamma)-\psi'(\frac{n}{2}+\gamma)-\chi_0(n,\gamma)^2
  \right\rbrace  + 
 \frac{1}{2 C_A f^\star (x_J,\mu_F)} \int_{x_J}^1 \de{}{\zeta} \Phi_{int}(\zeta,n,\gamma,\omega,J)
 \end{aligned}
\end{equation}

where we have
\begin{equation}
C_q=  \left( \frac{85}{18}+\frac{\pi^2}{2} \right) C_A-  \left( \frac{9}{2}+\frac{\pi^2}{3}-3 \log{2} \right) C_F -\frac{5}{9} N_F 
\end{equation}
\begin{equation}
C_g=   \left( \frac{1}{12}+\frac{\pi^2}{6} \right) C_A -\frac{1}{12} N_F 
\end{equation}

\begin{equation}
\begin{aligned}
 \Phi_{int}(\zeta,n,\gamma,\omega,J) =  -\frac{\zeta^{-\omega+\gamma-\frac{n}{2}}}{2 C_F \bar{\zeta}\zeta^2} 
  \left\lbrace \left[ 
2 C_A f_g \left(\frac{x_J}{\zeta},\mu_F\right) \left(
 C_A \left(1-\zeta \bar{\zeta} \right)^2 + N_F T_R \zeta \bar{\zeta}\left(\zeta^2 +\bar{\zeta}^2 \right) 
  \right)   +  \right. \right.  \\
\left. \left.  + C_F f_q\left( \frac{x_J}{\zeta},\mu_F\right) \left(
2 C_F \zeta \left(1+\zeta^2 \right)+C_A \left(1-2\zeta \right) \left(2-\zeta+\zeta^2 \right)
 \right)
   \right]  \right. \\
 \left.
 \left[
 B_{\zeta }\left(\frac{n}{2}+1-\gamma,0\right) +\zeta^n \left( 
 B_{\zeta }\left(-\frac{n}{2}+1-\gamma,0\right)+B_{\zeta }\left(\frac{n}{2}+\gamma,0\right)+B_{\zeta }\left(-\frac{n}{2}+\gamma,0\right)
 \right) \right]
 \right\rbrace + \\
 +\frac{2 \zeta^{-\omega}}{C_F \bar{\zeta} \zeta^2} f_g\left( \frac{x_J}{\zeta},\mu_F\right) \left\lbrace
 C_F N_F \zeta^2 \bar{\zeta}^2 + 2\log{\frac{k_J}{\mu_F}} \left[ 
C_A^2 \left(1-\zeta \bar{\zeta} \right)^2 + T_R C_F N_F \zeta \bar{\zeta} \left(\zeta^2 \bar{\zeta}^2 \right)
  \right] -      \right. \\ 
  \left. - 2 C_A \zeta^{2 \gamma} \log{\left(\bar{\zeta} R \right)} \left[
C_A \left(1-\zeta \bar{\zeta}  \right)^2  +N_F T_R  \zeta \bar{\zeta} \left(\zeta^2 \bar{\zeta}^2 \right)
   \right] + \right. \\
   \left. + \chi_0\left(n,\gamma \right) \left[  
   C_A \left(1-\zeta \bar{\zeta}  \right)^2\left(1+\zeta^{2\gamma} \right)+ N_F T_R  \zeta \bar{\zeta} \left(1-2 \zeta \bar{\zeta} \right) \left(C_F + \zeta^{2\gamma}C_A \right)
   \right]
 \right\rbrace + \\
  +\frac{ \zeta^{-\omega}}{ \bar{\zeta} \zeta^2} f_q\left( \frac{x_J}{\zeta},\mu_F \right)  \left\lbrace
 \zeta \bar{\zeta} \left( C_A \zeta +C_F \bar{\zeta} \right) + 
 2\log{\frac{k_J}{\mu_F} } 
 \left[ C_A \bar{\zeta}  \left(2-2 \zeta  +\zeta^2\right) +  C_F  \zeta  \left( 1+\zeta^2 \right)   \right] 
  \right.  -   \\ 
  \left.     
   - 2 C_F \zeta^{2 \gamma} \left(2-3\zeta \bar{\zeta} \right)  \log{\left(\bar{\zeta} R \right)} 
   + \chi_0\left(       n,\gamma    \right) 
    \left[  
   C_A \bar{\zeta} \left(2-2 \zeta +\zeta^2 \right)+ C_F \left(
   \zeta \left( 1+ \zeta^2 \right)+ \zeta^{2\gamma} \left( 2 -3 \zeta \bar{\zeta} \right)
    \right) 
   \right]
 \right\rbrace - \\
 -\frac{2}{C_F \bar{\zeta}} \left\lbrace
C_A^2 f_g\left( \frac{x}{\zeta},\mu_F  \right)  \left[   2\log{\frac{k_J}{R \mu_F}}+ \chi_0\left(n,\gamma \right)     \right] + \right. \\
\left.
+C_F f_q\left( \frac{x}{\zeta},\mu_F  \right)  \left[
C_F \left(1+\zeta^2 \right)\log{\frac{k_J}{R \mu_F}}+C_A \chi_0\left(n,\gamma \right)    +2 \left(C_F-C_A \right)\log{\bar{\zeta}}
\right]
\right\rbrace
 \end{aligned}
\end{equation}
where $\bar{\zeta}=1-\zeta$, $T_R=\frac{1}{2}$, $C_F=\frac{N_C^2 - 1}{2 N_C}$ and $C_A=N_C$, the number of colors. The incomplete Beta function is defined as $B_{\zeta}\left( a,b \right) = \int_0^\zeta \de{}{x} x^{a-1} \left(1-x \right)^{b-1}$.

\end{appendices}

\section{Bibliography}

\bibliography{veto_biblio.bib}

\bibliographystyle{ieeetr}

\end{document}